\documentclass[preprint,showpacs,superscriptaddress]{revtex4-1}

\usepackage{amsfonts}
\usepackage{amssymb}
\usepackage{amsmath}
\usepackage{graphicx}
\usepackage{soul}
\usepackage[table,xcdraw]{xcolor}
\usepackage{dcolumn}
\usepackage{bm}
\usepackage[T1]{fontenc}

\definecolor{Gray}{gray}{0.89}

\begin{document}

\title{Thermal relaxation in titanium nanowires: signatures of inelastic electron-boundary scattering in heat transfer}

\author{Teemu Elo}
\affiliation{Low Temperature Laboratory, Department of Applied Physics, Aalto University, Espoo, Finland}
\author{Pasi L\"ahteenm\"aki}
\affiliation{Low Temperature Laboratory, Department of Applied Physics, Aalto University, Espoo, Finland}
\author{Dmitri Golubev}
\affiliation{Low Temperature Laboratory, Department of Applied Physics, Aalto University, Espoo, Finland}
\author{Alexander Savin}
\affiliation{Low Temperature Laboratory, Department of Applied Physics, Aalto University, Espoo, Finland}
\author{Konstantin Arutyunov}
\affiliation{National Research University Higher School of Economics, Moscow Institute of Electronics and Mathematics Moscow, Russia}
\affiliation{Kapitza Institute for Physical Problems, Russian Academy of Science, Moscow, Russia}
\author{Pertti Hakonen}
\affiliation{Low Temperature Laboratory, Department of Applied Physics, Aalto University, Espoo, Finland}

\begin{abstract}
We have employed noise thermometry for investigations of
thermal relaxation between the electrons and the substrate in
nanowires patterned from 40-nm-thick titanium film on top of silicon wafers
covered by a native oxide. By controlling the electronic temperature $T_e$ by Joule
heating at the base temperature of a dilution refrigerator, we probe the electron-phonon coupling and the thermal boundary resistance at temperatures $T_e= 0.5 - 3$ Kelvin.
Using a regular $T^5$-dependent electron-phonon coupling of clean metals and
a $T^4$-dependent interfacial heat flow, we deduce a small contribution for the direct
energy transfer from the titanium electrons to the substrate phonons due to inelastic electron-boundary scattering.

\end{abstract}


\maketitle

\section{Introduction}

Thermal properties of nanoscale structures and their particular energy relaxation
processes form a key issue for various practical applications ranging from
quantum computing to ultra sensitive radiation detectors and quantum-limited
calorimeters. By providing the major mechanism
for heat transfer between electrons and phonons at typical cryogenic temperatures \cite{Lounasmaa1974}, the inelastic electron-phonon  scattering
determines the effective temperature of the electron sub-system, and it plays
an essential role in the operation of different low-temperature micro and nano devices.
Optimization, or preferably tuning, of this parameter in a wide range sets the
basis for successful development and improvement of sensitivity of a vast class of
devices.

Thermal relaxation of electrons in  metallic cryogenic nanodevices takes place typically either by
electronic diffusion or by electron -- phonon coupling (e-ph). The e-ph relaxation processes lead to a temperature-dependent
thermal conductance which varies with the temperature $T$ in a power law fashion $%
\propto T^{n}\Delta T$ where $n=3-5$ \cite{Giazotto2006} and $\Delta T$ denotes the temperature difference between the electrons and phonons. The heat flow due to the e-ph coupling may lead to an overheated state of the phonons in the metallic conductor owing to the Kapitza resistance between the phonon systems of the dielectric substrate and the metal. In addition to the regular Kapitza resistance $\propto 1/T^3$, there may also be a direct electronic heat transfer channel from the electrons to the substrate phonons  \cite{Hubermann1994,Sergeev1998,Mahan2009,Low2012}. In clean systems, this direct coupling facilitated by the inelastic electron-boundary scattering has the same form as the Kapitza-resistance-limited heat conductance, in which the phonon temperature of the hot side is replaced by the electron temperature. The relative strength of these two parallel heat transfer channels across the film-dielectric interface is an open question, and this issue forms the main objective of the present study.

We have chosen thin, narrow titanium wires for our studies. Titanium metal has been successfully employed
in various ultrasensitive hot-electron bolometers both in the superconducting and normal states \cite{Wei2008,Karasik2011}.
In spite of the quite intensive experimental studies of
thermal relaxation in titanium micro and nanostructures \cite%
{Wu1998,Hsu1999,Manninen1999,Gershenson2001,Horn2003,Fukuda2007,Karasik2007,Wei2008,Day2008,Karasik2009}%
, the reported results for titanium and its alloys distinctly vary both in their absolute
magnitude as well as in their functional form
\cite{Karvonen2005,Karasik2011}.

We study titanium nanowires having cross sections, thickness $\times$ width $=d \times w$, of about $40\times 38\ldots 50$ nm$%
^{2}$, and find the best-fit exponent in the electron -- subsrate phonon heat conductance law to be close to $%
n=3$. Such an observation is consistent with the predictions for the
electron -- phonon coupling in disordered metallic films as well as with
the behavior governed by Kapitza resistance. Our analysis of the electron relaxation and thermal balance in Ti wires is based on a model that includes the overheated state of the phonons with an intermediate temperature in the film as well as the direct energy transfer from the electrons to the substrate phonons via inelastic electron scattering at the film-substrate boundary. Our model allows for investigations of the cross-over behavior between the distinct power law regimes which are dominated by different  relaxation processes. Using such a  model with realistic materials parameters, we obtain evidence for a weak contribution of the inelastic electron-boundary scattering. Our result is obtained at temperatures between $1$ K
and $3$ K in a regime where the Kapitza resistance between the film and the substrate phonons reduces the major heat flow mediated by the regular electron-phonon coupling. However, as this Kapitza resistance is poorly known, we also discuss other possible interpretations of our data, such as a modification of the electron-phonon coupling due to disorder, yielding a different value for the exponent $n$.

\section{Thermometry and heat transfer}

Our experiment relies on the sensitivity of noise thermometry of  electrons in nanoscale samples \cite{Wu2010,Chaste2010,Santavicca2010}. Noise thermometry in quasi-equilibrium is possible when the electron-electron scattering is strong and a local, spatially varying temperature profile $T(x)$ can be defined \cite{Blanter2000}. The average effective temperature is related to the bias voltage $V$ via  $T_e={FeV}/2k_{B}$ where the Fano factor $F\equiv \frac{S_{I}}{2eI}$ is given by the current noise power spectrum $S_I$ and  the current through the sample is $I$. Using the diffusion constant $D=3.8 \times 10^{-4}$ m$^2$/s obtained from the electron mean free path of 2.5 nm and the Fermi velocity of 300 km/s \cite{Sanborn1989}, we obtain for the electron-electron (e-e) interaction length $l_{e-e} = 0.36$ $\mu$m at $T_e=0.1$ K (the worst case estimate corresponding to the maximum length). As $l_{e-e} \ll L$, we are in the strong electron-electron interaction regime  (the hot electron regime), and the Fano factor becomes $F=\sqrt{3}/4$. The value of $F$ remains constant till the inelastic processes start to remove energy from the electronic system. The proportionality given by the Fano factor between the effective temperature $T_e$ and the bias $V$ is also valid  in the high-bias regime where the electron-phonon scattering dominates the heat transfer \cite{Wu2010}. This is because the noise power and the related $T_e$ are connected by an effective Johnson noise relation.
 Recently, noise thermometry
has been utilized in the studies of novel nanoscale
systems when the implementation of regular thermometry is problematic \cite%
{Tarkiainen2003,Wu2010,Chaste2010,Santavicca2010,Fong2012,Laitinen2014,McKitterick2014}.

\begin{figure}[tbp]
\centering
\includegraphics[width=0.7\textwidth]{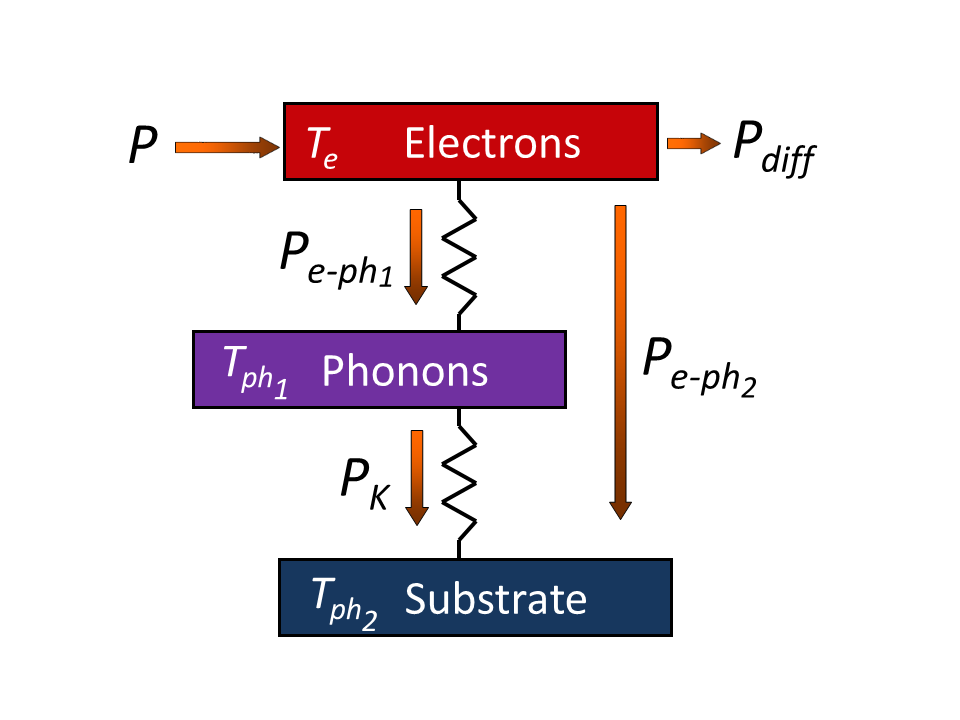}
\caption{Simplified thermal model of a conducting nanostructure on an
isolating substrate. $P=I^{2}R_{n}$ denotes the Joule heating by the current $I$ passing through
the resistance $R_{n}$  of a normal wire, and $P_{\mathit{diff}}$ specifies the heat
carried away by electronic diffusion. The heat flows $P_{e-ph_1}$, $P_{K}$, and $P_{e-ph_2}$
are defined in Eqs. \protect\ref{power_ep}, \protect\ref{Kapitza_tot}, and \protect\ref{eKapitza},
respectively. }
\label{thermal_model}
\end{figure}

The Joule heating power, $P=I^{2}R$ in a wire with resistance $R$, can be dissipated into phonons via the electron--phonon
(e-ph) interaction. The poor coupling of the film phonons with the substrate phonons will affect the
e-ph scattering rates and, consequently, influence the electron temperature once the e-ph scattering becomes stronger than the e-e scattering. Here,
we employ the  energy balance model for the \emph{sample electron --
sample phonon -- substrate phonon} coupling \cite%
{Bron1977,Roukes1985,Wellstood1994,Bergmann1990} schematically presented in
Fig.~\ref{thermal_model}, including also the inelastic electron-boundary scattering that couples electrons directly to the substrate phonons \cite{Hubermann1994,Sergeev1998}.
In this model, the ratio of $P_{diff}$ to $P_{e-ph}=P_{e-ph_1} + P_{e-ph_2}$
determines the crossover from the nearly parabolic temperature profile in the hot electron regime to a more constant profile in the case of strong electron-phonon scattering. In our analysis of the significant part of the wire (see Sect IIIB), we assume $P_{diff} \ll P_{e-ph}$, which is valid only in the strong bias regime.
The temperature dependence $P_{diff}\propto T_e^{2}$ can be obtained from the
Wiedemann-Franz law utilizing the measured resistance of the wire.

Apart form the electronic heat
diffusion, $T_{e}$ is determined by the three thermal
resistances responsible for (a) the heat exchange between the electrons and phonons
in the conductor ($ P_{e-ph_1}$ in Fig. 1), (b) the acoustic coupling between the
phonons in the metallic conductor and the phonons in the insulating substrate
($ P_{K}=P_{ph_1-ph_2}$), and (c) the coupling between
the electrons in the metallic conductor and the phonons in the insulating substrate ($ P_{e-ph_2}$).
When $P \gg P_{\mathit{diff}}$, the power-law exponent (see below) is determined by the dominating thermal resistance
along the main heat flow path.

In regular
metals, the power flow between the electron and phonon systems, corresponding to  $P_{e-ph_1}$ in Fig. 1 can be
expressed as follows \cite{Kaganov1957,Allen1987,Giazotto2006}:

\begin{equation}
P_{e-ph_1}=\Sigma _{e-ph_1}\Omega (T_{e}^{p+2}-T_{ph}^{p+2}),  \label{power_ep}
\end{equation}%
where $T_{ph}$ denotes the phonon temperature, $\mathit{\Sigma _{e-ph_1}}$ is
a material-dependent coupling coefficient, $\Omega $ is the volume of the sample,
and $p+2$ is the exponent specified by the electron-phonon scattering rate $%
1/\tau _{e-ph_1}\propto T^{p}$. Nowadays, it is commonly accepted that, in clean
metals, the electron-phonon scattering rate is proportional to \textit{T${}^{p}$}
with $p=3$ \cite{Roukes1985,Wellstood1994,Sergeev2000}. Similar behavior is also observed  in thin metallic films \cite{Giazotto2006}, including titanium. In our analysis, we take $\Sigma _{e-ph_1}=1.3 \times 10^9$ Wm$^{-3}$K$^{-5}$ from Ref. \onlinecite{Manninen1999}.

The physics of the e-ph coupling becomes more complicated in dirty metals \cite%
{Sergeev2000} or in systems with a restricted geometry where the electron-phonon
scattering or/and the phonon spectrum \cite{Karvonen2007} can significantly be
modified compared to the bulk materials in the clean limit. In disordered systems,
the electron-phonon scattering rate follows \textit{T${}^{2}$} or the \textit{T$%
{}^{4}$} ($p=2$ or 4) law depending on the scattering mechanisms \cite%
{Sergeev2000}. Evidence of $p=2$ in disordered films has been obtained, for example, in Nb and Ti \cite{Gershenzon1990, Wei2008}.

In clean metals, the
dimensionality of the sample may lead to a modification of the electron-phonon relaxation
time due to a reduction in the available phonon modes. According to
the Debye model \cite{Kittel}, the change of the acoustic phonon spectrum from 3D to 2D
leads to a weaker temperature dependence of the electron-phonon power flow,
namely to a $T^{d+2}$ law where $d$ is the phonon dimensionality \cite{DiTusa1992}.
Evidence of this behavior has been obtained in Ref. \onlinecite{Karvonen2007}.

The Kapitza thermal boundary resistance originates
from the mismatch of acoustic properties across an interfacial boundary, leading to a phonon
temperature difference between the two materials $T_{ph_{1}}\neq T_{ph_{2}}$%
. The corresponding power flow through the interface can be expressed as
\cite{Pollack1969,Lounasmaa1974}:
\begin{equation}
P_{K}=\frac{1}{4} A_{K}S(T_{ph_{1}}^{4}-T_{ph_{2}}^{4}),  \label{Kapitza_tot}
\end{equation}
where the coupling coefficient has been expressed
in terms of the linearized Kapitza conductance of the interface between the
two materials $G_{K}=A_{K}T^{3}$ and  the area of the interface $S$.
Typically, the coefficient $A_{K}$ varies between $100$ and $1000$ Wm${}^{-2}$K$%
{}^{-4}$ for the common metal-to-dielectric interfaces used in microfabricated
systems \cite{Swartz1989}.
The standard value for the Kapitza conductance between a metallic film and the common dielectrics amounts to $A_K^0 =500$ Wm${}^{-2}$K$^{-4}$ \cite{Roukes1985,TaskinenPhD}. In our wire geometry  with a nearly square-shaped cross section, we have adopted the Kapitza conductance $A_K = A_K^0/2$, where the factor of two approximates the reduction of the effective phonon modes in the  wire geometry. The reduction value reflects ray-tracing of phonons, where the propagation angles exceeding $\sim 45^\circ$ cannot fully escape the wire without collisions with the edges. Even a larger loss of coupling would be obtained if van Hove singularities in the phonon density of states are taken in to account, in a similar fashion as is predicted for the electron-phonon coupling in reduced dimensions \cite{Bilotsky2008}.

The inelastic electron scattering at the interface between a metallic film and an
insulating substrate may provide an additional channel for the energy transfer
from electrons to the phonons of the substrate \cite{Hubermann1994,Sergeev1998}, proportional to $T_e-T_{ph_2}$.
At low temperatures near equilibrium, the contribution $P_{e-ph_2}= \frac{1}{4}S A_{K}^{e}(T_e^4-T_{ph_2}^4)$ to the interfacial "electronic Kapitza" thermal conductance is
given by \cite{Sergeev1998}
\begin{equation}
A_{K}^{e}=\frac{3\pi \hbar }{35\zeta (3)k_{B}}\frac{\gamma u_{l}}{\tau _{e-ph_1} T_e^3%
}\left[ 1+2\left( \frac{u_{l}}{u_{t}}\right) ^{3}\right] ,  \label{eKapitza}
\end{equation}%
where $u_{l}$ and $u_{t}$ are the longitudinal and transverse sound
velocities, respectively, and $\gamma $ is the Sommerfeld constant. If in a
metal $\tau _{e-ph_1}\propto 1/T^{3}$, then this inelastic electron-boundary scattering contribution to the interfacial thermal
conductance has the same temperature dependence as the regular acoustic phonon-phonon Kapitza term. The
mechanism can be significant in those metallic films which have a strong electron-phonon
coupling. The estimations for our samples based on Eq. \ref{eKapitza} using $%
\tau _{e-ph_1}\propto T^{-3}$ from Ref. \onlinecite{Manninen1999} yield $%
A_{K}^{e}=300$ to $400$ Wm$^{-2}$K$^{-4}$.
However, since Eq. \ref{eKapitza} has been derived
ignoring the reduction of the electron-phonon coupling near the interface, we regard $A_{K}^{e}$ as a poorly known parameter and expect its actual magnitude to be smaller, possibly a few times smaller, than evaluated from the above equation.

\begin{figure}[tbp]
\centering
\includegraphics[width=0.7\textwidth]{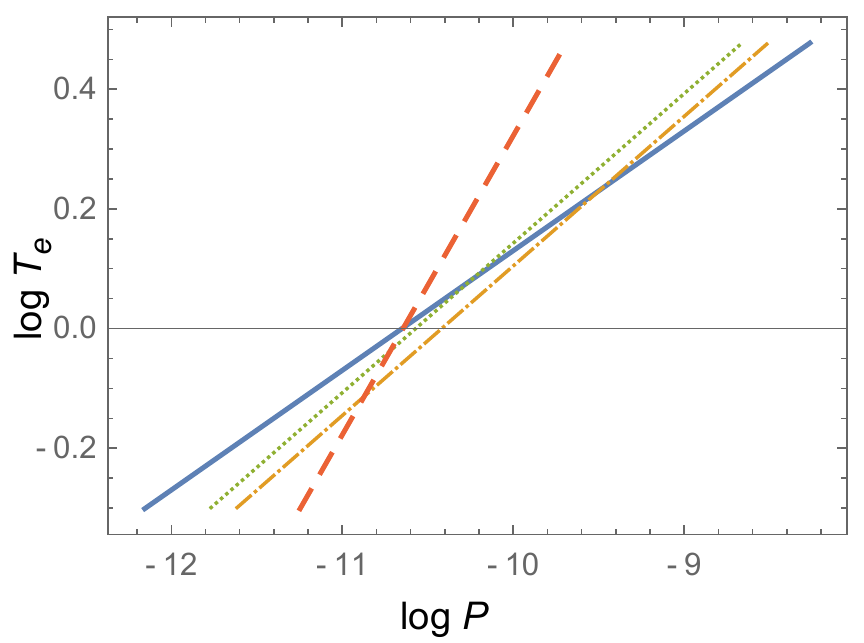}
\caption{ Comparison of the heat flow terms on a $\log P$ vs $\log T_e$ plot:  $ P_{e-ph_1}$  (solid), $ P_{e-ph_2}$ (dash-dotted),  $P_{K}$ (dotted), and $P_{diff}$ (dashed). For the employed parameter values, see text; $P$ and $T_e$ are expressed in terms of W and K, respectively.}
\label{terms}
\end{figure}
The different heat flow terms for a Ti wire with a cross section of $d \times w = 40 \times 43$ nm$^2$ and having a length of $L = 10$ $\mu$m are illustrated in Fig. \ref{terms}. In addition to the contributions $ P_{e-ph_1}$, $ P_{e-ph_2}$, and  $P_{K}$, we depict also the outdiffusion of heat $P_{diff}$; the large ratio of $L/d$ excludes any substantial phonon conductance along the wire. According to the comparison, we expect initially a regime governed by the electronic heat diffusion, which crosses over to the Kapitza resistance limited regime above $\sim 1$ K. The direct coupling of electrons to the substrate phonons given by the theoretical formula in Eq. \ref{eKapitza} is indicated by the dashed-dotted line.   In our analysis, we use literature values for the term $ P_{e-ph_1}$ \cite{Manninen1999} while for  $P_{K}$ we employ a semiempirical value based on Refs. \onlinecite{Roukes1985} and \onlinecite{Bilotsky2008}, which leaves only the magnitude of $ P_{e-ph_2}$ as a parameter to fit to the data.

\section{Thermal modeling}
\subsection{Two temperature model: Hot phonons in the wire}
To clarify our thermal model based on Fig. 1, we consider a long, voltage-biased normal wire coupled to the films of the same thickness on both sides.
We assume that the electronic thermal diffusion of heat  $P_{diff}$ can be neglected at high bias, i.e., the electronic temperature inside the wire, $T_e$, is constant as $P_{e-ph_1}$ dominates the heat conduction. The Joule heat generated
in the wire per unit volume, $\frac{I^2}{\sigma S^2}$, is first transferred to the phonon heat bath inside the wire, which is characterized
by the temperature $T_{ph_1}$,
\begin{eqnarray}
\frac{I^2}{\sigma w^2d^2} - \Sigma_{e-ph_1} (T_e^5 -T_{ph_1}^5) = 0,
\label{heating1}
\end{eqnarray}
where $\sigma$ is the conductivity. At the second stage, the phonons are cooled via the Kapitza conductance mechanism
\begin{eqnarray}
d \Sigma_{e-ph_1} (T_e^5 -T_{ph_1}^5) = \frac{1}{4} A_K (T_{ph_1}^4 -T_{ph_2}^4).
\label{heating2}
\end{eqnarray}
Here $T_{ph_2}$ is the base temperature.
Using Eqs. \ref{heating1} and \ref{heating2}, we can solve
for the electronic temperature
\begin{eqnarray}
T_e = \left(\frac{P}{\Sigma_{e-ph_1}} + \left(T_{ph_2}^4 + \frac{dP}{\frac{1}{4} A_K}\right)^{5/4}\right)^{1/5}.
\label{modelT}
\end{eqnarray}
This formula is fitted to our data in Fig. \ref{electron_T}. The role of the direct electron-phonon contribution $P_{e-ph_2}$ is to lower the heat flow $P \rightarrow P-P_{e-ph_2}$ in Eq. \ref{modelT}. For simplicity, we employ the ratio $A_{K}^{e}/(A_{K}+A_{K}^{e})$ for the reduction factor which is valid in the high-temperature limit.

\subsection{Analysis of thermal gradients} \label{LTA}

\begin{figure}[!ht]
\includegraphics[width=10cm]{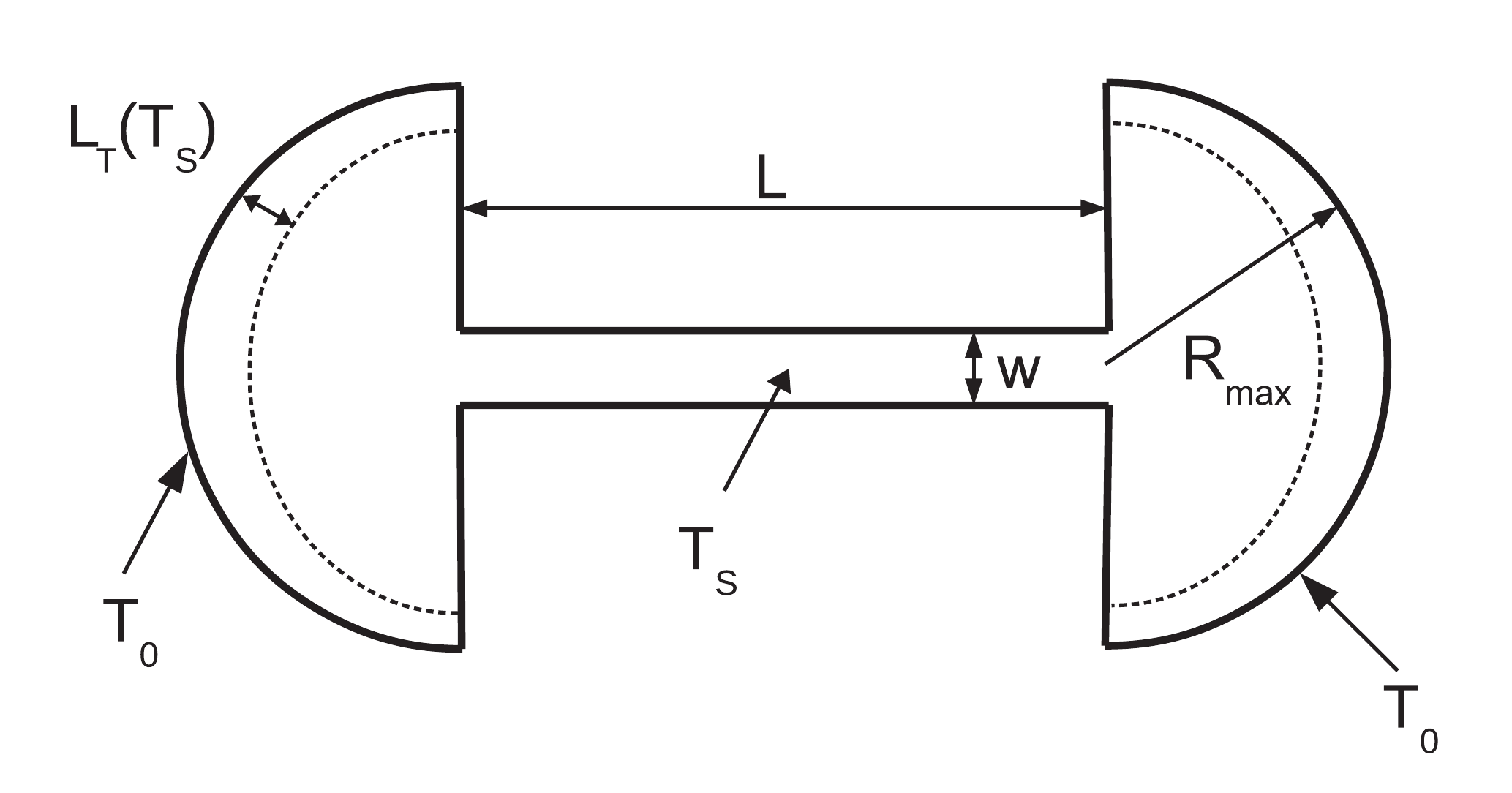}
\caption{Geometry of the nanowire between two wide leads used for the analysis of the relaxation of $T_e$ outside the narrow section having a width $w$ and length $L$. For the definition of the other symbols, see text.}
\label{fig:wire}
\end{figure}

Besides the assumption $T_e=\textrm{const}$ in the narrow wire part, our analysis assumes that the scale of the relaxation of $T_e$ in the overheated wide sections of the leads is so short that the effective thermal noise of the leads can be neglected.  To justify this assumption, we derive a characteristic length scale for the temperature drop at the ends of the wire. We consider a normal wire coupled to the two wide normal leads on both sides as depicted in Fig. \ref{fig:wire}.
The wire and the leads are made of the same film thickness $d$. We assume that at some large distance, $R_{\max}$, from the wire ends
the temperature of the leads becomes equal to the temperature of the substrate $T_{ph_2} \simeq 20$ mK.

The temperature both in the wire and in the leads can be found from the equation
\begin{eqnarray}
\frac{\pi^2 k_B^2 \sigma d}{6e^2}\nabla^2 T_e^2 + \frac{j^2 d}{\sigma} - \Sigma^{tot}_{e-ph_2} d (T_e^4 - T_{ph_2}^4) = 0,
\label{lead}
\end{eqnarray}
where we have approximated the experimentally determined heat flow  by the formula $\Sigma^{tot}_{e-ph_2}  (T_e^4 - T_{ph_2}^4)$ per unit volume (see Sect. \ref{res}),
and $j$ is the current density. Inside the wire $j=I/wh$ while
$j = \frac{I}{\pi h r}$ in the lead at a distance $r$ from the end of the wire.
At a sufficiently high bias, the temperature in the leads in the vicinity of the wire ends equals to
\begin{eqnarray}
T_S(r) = \left( T_{ph_2}^4 + \frac{j(r)^2}{\sigma\Sigma^d_{e-ph_2}} \right)^{1/4}.
\end{eqnarray}
At some distance  $r\sim R_{\max}$ the temperature drops to the equilibrium value $T_0$. The characteristic length
scale at which this happens is $L_T(T_S)$, where
\begin{eqnarray}
L_T(T)=\sqrt{\frac{\pi^2 k_B^2\sigma}{6e^2\Sigma^{tot}_{e-ph_2} T^{2}}}
\end{eqnarray}
is the temperature relaxation length. Since the resistance of the wire,  $R_{\rm wire}=L/\sigma wd$, is much higher than
the resistance of the leads, $R_{\rm lead}=\ln(R/w)/\pi \sigma d$, the noise in this regime may be approximated as
$S=\frac{4 k_B T_S}{R_{\rm wire}}$, i.e. the noise from the end sections can be neglected. This analysis is consistent if $R_{\max} > L_T(T_S)$, which is satisfied at a sufficiently large Joule heating.
The values of the relaxation length at $T=1$ K, which is a typical value of the temperature $T_S$ in the experiment, are listed in Table 1.

\section{Experimental procedures}

\begin{figure}[tbp]
\includegraphics[width=12cm]{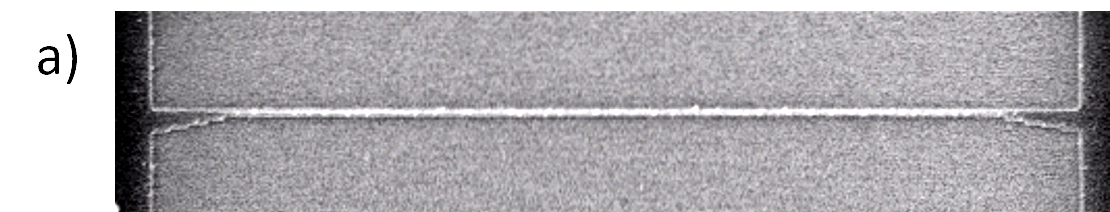} %
\includegraphics[width=14cm]{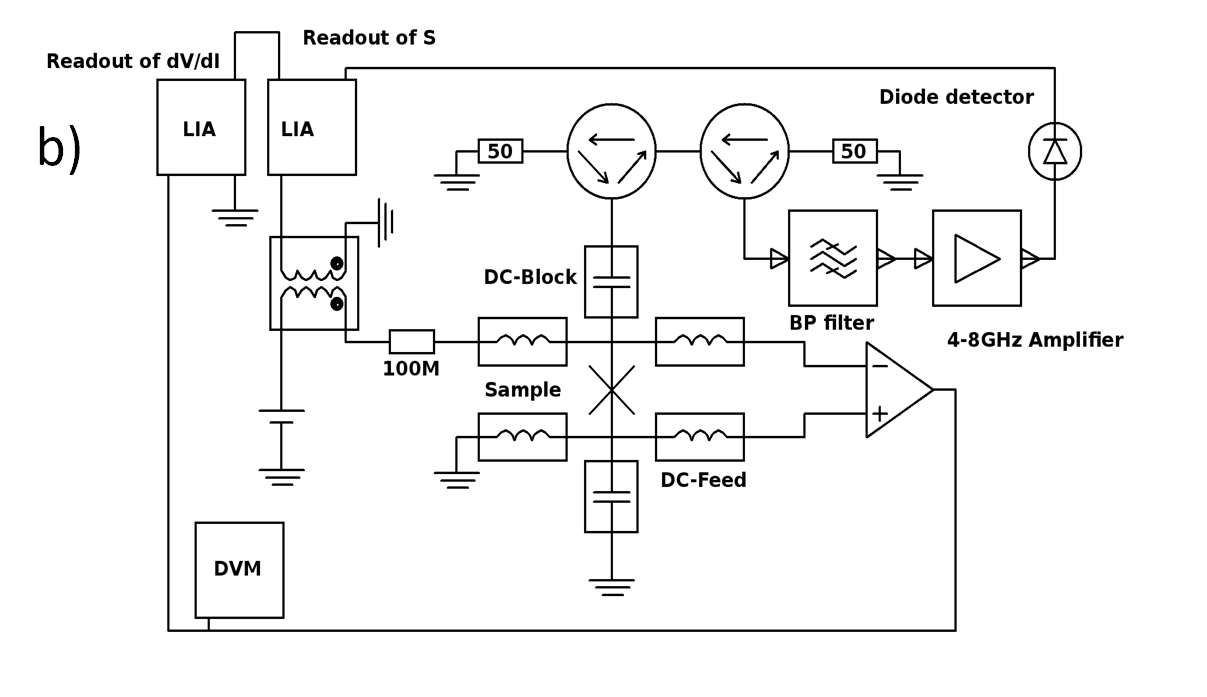}
\caption{a) SEM image of a typical titanium nanowire;  the distance between the "black" reservoirs is $L = 10$ $\mu$m.
b) Schematic diagram of the measurement circuit. LIA denotes an audio-frequency lock-in amplifier,
DVM stands for a digital voltmeter. }
\label{sample}
\end{figure}

Our Ti wires were evaporated from a 99.999\%
titanium source in a high-quality e-gun evaporator at pressures below $10^{-9}$ mBar.
High-resistivity silicon wafers ($\sim 10$ k$\Omega$cm) with a natural oxide layer were used as the insulating substrate. The
samples were fabricated using a lift-off process based on e-beam lithography and a PMMA resist
mask. Fig. \ref{sample}a displays an SEM image of a typical structure.
Sample B (see Table I) was measured without any additional processing steps.
Samples A and C were cleaned using $Ar^{+}$ ion milling to reduce the cross
section of the nanowire \cite{Zgirski2008}. As a byproduct of the ion
milling process, the sample surfaces became efficiently polished down to $\pm $ 1 nm scales.
The basic
parameters of the measured samples are specified in Table I. The bonding pads on the sample chip were $100 \times 100$ $\mu$m$^2$.

\begin{table}[tbp]
\begin{tabular}{|c|c|c|c|c|c|c|c|c|}
\hline
& Dimensions nm$^3$ & \textit{R${}_{n}$} & \textit{I${}_{c}$ } & $L_T(1K)$ & $T_c$  &
treatment \\ \hline
A & 50 $\times$ 40 $\times$ 10,000 & 4 k$\Omega$ & 10 nA & 2.7 $\mu$m & 0.25 K & sputtered \\ \hline
B & 43 $\times$ 40 $\times$ 10,000 & 7 k$\Omega$ & 7 nA & 2.7 $\mu$m & 0.10 K &  not etched \\ \hline
C & 38 $\times$ 40 $\times$ 10,000 & 5 k$\Omega$ & 9 nA & 3.7 $\mu$m & 0.16 K &  sputtered \\ \hline
\end{tabular}%
\caption{Parameters of the Ti nanowires labeled as A, B, and C:
\textit{R${}_{n}$} denotes the normal state resistance of the wire, $I_{c}$ is the
superconducting critical current, and $L_T(\textrm{1~K})$ denotes the temperature dependent thermal relaxation length (see Sect. \ref{LTA}) in the wide leads evaluated at 1~K using the measured electron-substrate phonon coupling given in Table II.
The superconducting transition temperature $T_{c}$ is determined from the mid-point of the transition
 and the last column
specifies the post treatment operations. }
\end{table}

The length of our wires was chosen to be quite long (10 $\mu $m) in order to
alleviate the influence of the temperature gradients at the ends of the wires on
the analysis and to reduce the influence of the longitudinal electronic heat
diffusion along the wire. The samples were mounted on a copper sample holder
in a Bluefors LD250 ${}^{3}$He/${}^{4}$He dilution refrigerator with a base
temperature of 10 mK. The heating power applied to the electron system of the
nanowire was calculated from the applied electrical current and the wire
resistance. The effective noise temperature scale was calibrated against the shot noise of
an aluminum oxide tunnel junction; the scale was verified against the effective thermal noise in the hot electron regime.

The electric measurement setup is schematically presented in Fig. \ref%
{sample}b. By employing both low and high frequency coupling components, the
setup facilitates audio-frequency conductance measurements and, simultaneously, a
recording of current fluctuations at GHz-frequency.  Two bias-Ts and two inductors are connected
to the sample in order to measure the low-frequency AC conductance with a
lock-in amplifier (LIA) in a four probe configuration. Two coupling
capacitors are connected to the ends of the sample in order to couple the high-frequency current fluctuations to the 50 Ohm measurement cables. Good coupling of the sample noise to the 50 $\Omega$ system was ensured by using short bond wires and by avoiding capacitive shunting in the pad and lead structures on the chip. Two 4-8 GHz
circulators in series and a corresponding band pass filter (with a 0.5-dB pass band loss) are employed to
block the backaction noise of the cryogenic HEMT amplifier (Low Noise Factory,
LNF-LNC4\_8A) from reaching the sample, which is of particular importance
when checking for superconducting properties of nanoscale systems. The detection of noise over the spectrometer bandwidth of $BW \simeq 2$ GHz \footnote{Due to the bandpass filter in the output line and the frequency dependence of attenuation, the effective band width as referred to the maximum gain is reduced by a factor of two.} is performed using a
square-law diode detector (Krytar 203BK Schottky diode). A second lock-in amplifier
is recording differential current-induced noise $dS_{I}/dI$ which can be integrated to
yield the excess current noise $S_{I}(I)-S_{(}0)=\int_{0}^{I}(dS_{I}/dI)dI$ at
current $I$.  The operation
with AC-current modulation reduces possible drift in the gain of the HEMT
amplifier which easily hampers the performance of a noise spectrometer
\cite{Nieminen2016}.

\begin{figure}[tbp]
\includegraphics[width=0.48\textwidth]{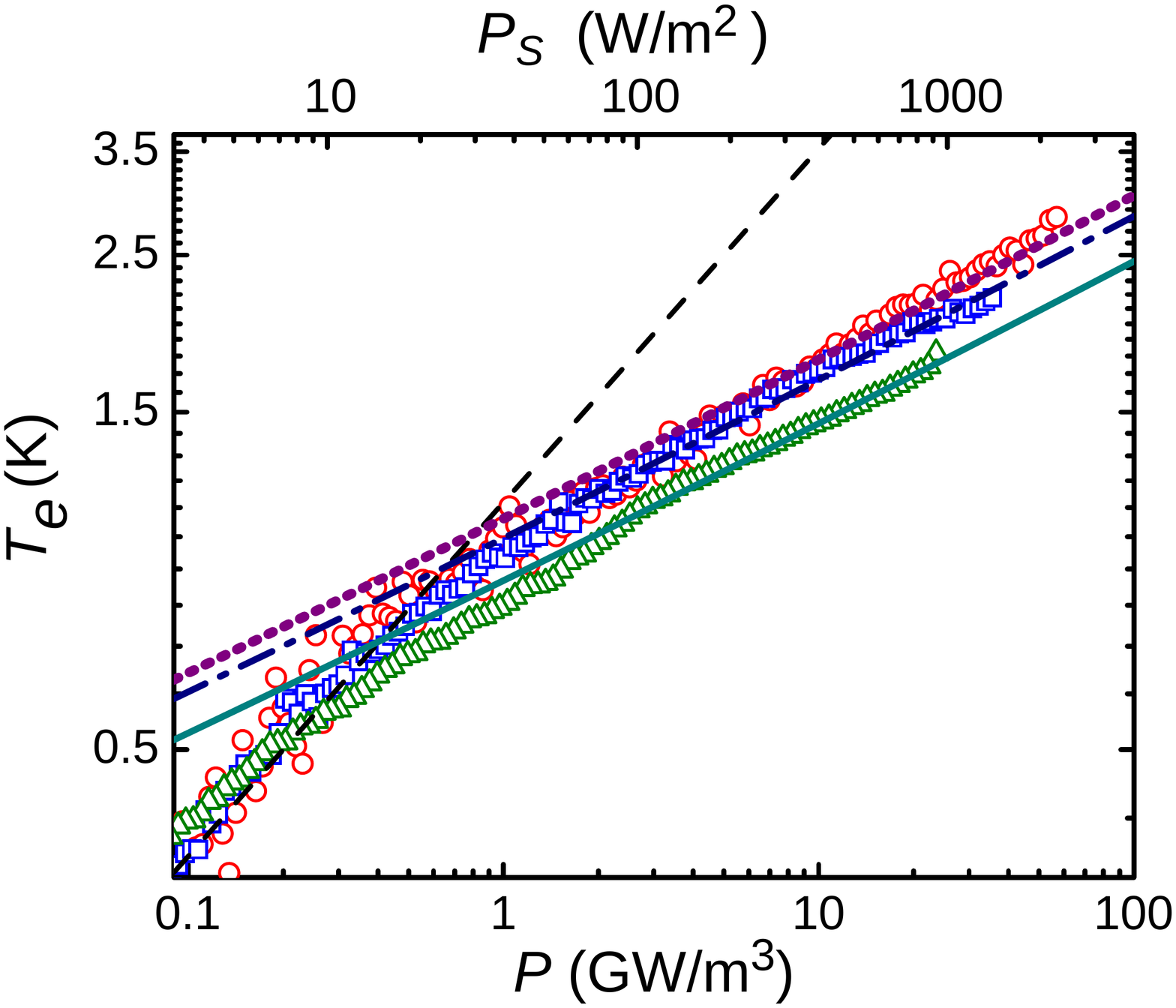}
\includegraphics[width=0.48\textwidth]{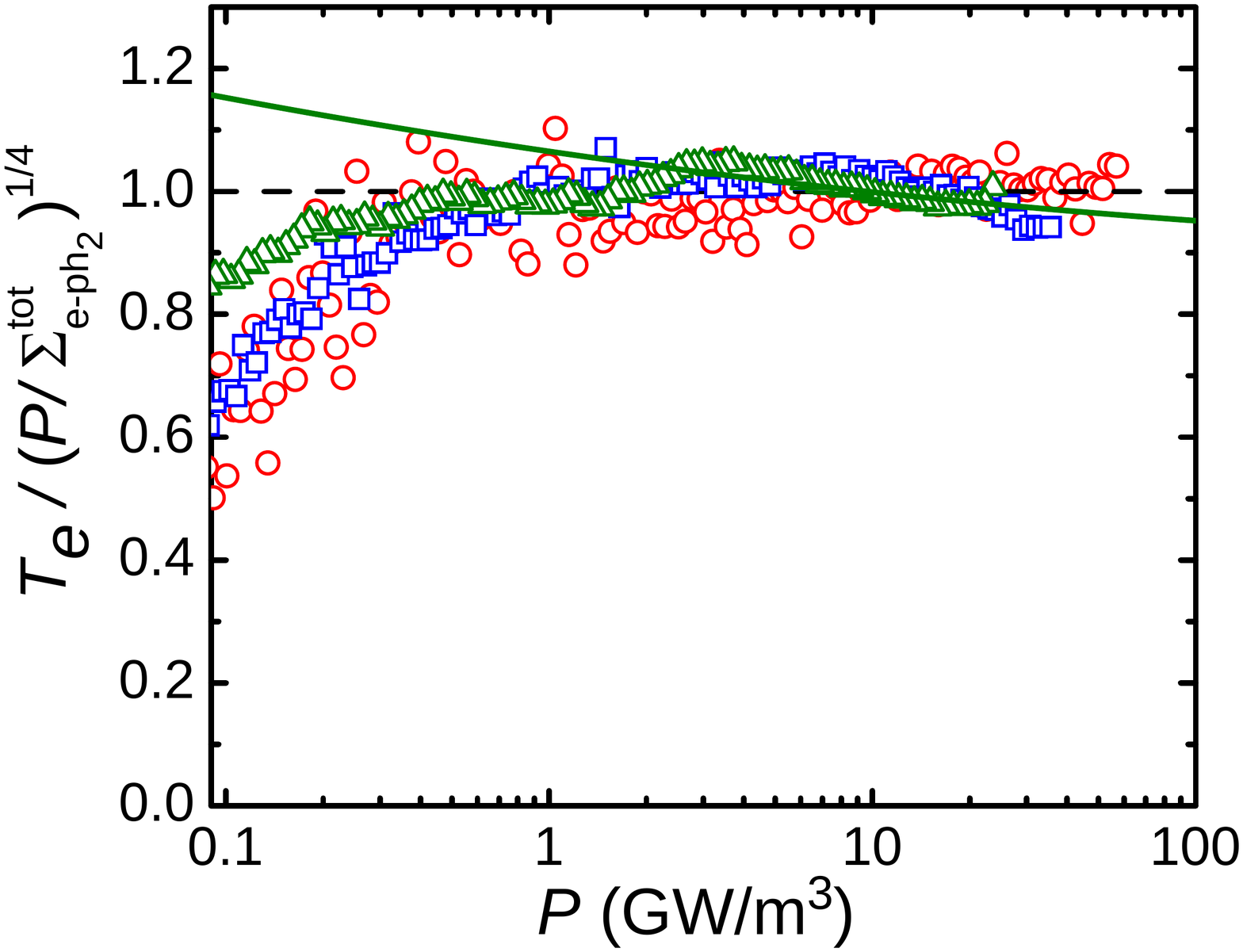}
\caption{a) Normal state electron temperature in Ti nanowires as a function
of the applied power density (bottom scale) and the power per contact area $P_S$ (top
scale) on a log-log scale. The dashed line $T_e \propto P^{1/2}$ denotes the calculated behavior for a hot electron system.
The dotted curve displays the behavior obtained from Eq. \ref{modelT}. The dash-dotted and solid curves are obtained from Eq. \ref{modelT} assuming
 that a fraction $A_{K}^{e}/(A_{K}+A_{K}^{e}) =0.25$ and 0.60 of the Joule heating, respectively, is
 conveyed to the substrate phonons directly via the inelastic electron-boundary scattering; the parameter
 values for $A_K^{e}$ are given in Table II. b) The measured electronic temperature
 divided by the heat flow scaled by the inverse $T^{4}
$ power law exponent. The solid curve corresponds to the fitted solid curve in Fig. \ref{electron_T}a;
the deviation at low bias is due to the electric heat diffusion along the wire.
The wire widths are denoted by: ($\Box$) -- 43 nm, ($\triangle$) -- 50 nm, and ($\circ$) -- 38 nm.
 }
\label{electron_T}
\end{figure}

\section{Results}\label{res}

Fig. \ref{electron_T}a displays the dependence of the effective electron temperature in
the titanium nanowire as function of the applied power density in W/m$^{3}$, and also
as a function of the power per contact area W/m$^{2}$ (top axis). We display our data at $T_e > 0.5 $ K only, since  the behaviour
of $T_{e}(P)$ below $0.5$ K was dominated by the electronic heat diffusion and it was influenced by
the superconducting
transition, the critical temperature of which slightly varied between the Ti samples
(see Table I). At $T_{e}>1$ K, the temperature dependence in each sample demonstrates a nearly
linear behaviour on the log-log scale. The dashed line in Fig. \ref{electron_T} with $T_e \propto P^{1/2}$ indicates the behavior of the hot electron regime with $ \langle T_e \rangle = FeV/2k_B$ with $F=\sqrt{3}/4$. The data agree quite well with the hot electron behavior of $P \propto T_e^2$ at bias voltages $V \lesssim 0.4$ mV, which also lends support to the correctness of the employed effective noise temperature scale.

 Three curves based on Eq. \ref{modelT} are displayed in Fig. \ref{electron_T}a, two with and one without the heat flow contribution due to $P_{e-ph_2}$. A better agreement is obtained with the term $P_{e-ph_2}$ included. The obtained values for the conductance due to the inelastic electron-boundary scattering \cite{Hubermann1994,Sergeev1998} are given in Table II. On the average, the obtained value $\langle A_K^e \rangle  = 80$ Wm$^{-2}$K$^{-4}$ is nearly by  an order of magnitude smaller than that evaluated from Eq. \ref{eKapitza}.

\begin{table}[tbp]
\begin{tabular}{|c|c|c|c|c|c|c|}
\hline
 & Width (nm) & $A_K^T$ (Wm$^{-2}$K$^{-4}$) & $A_K^e$ (Wm$^{-2}$K$^{-4}$) & $\Sigma_{e-ph_2}^{tot}$  (Wm$^{-3}$K$^{-4}$) \\ \hline
A & 50 & 340 & 200 & $2.1 \times 10^9$ \\ \hline
B & 43 & 220 & 55 & $1.4 \times 10^9$  \\ \hline
C & 38 & 180 & 0 & $1.1 \times 10^9$  \\ \hline
\end{tabular}%
\caption{Results derived from the thermal model for samples A, B, and C. Columns:
The width of the wire, $A_{K}^T=A_{K} + A_{K}^e$ is the total Kapitza conductance as
determined at an infinite electron--phonon coupling, $A_K^e$ is the extracted boundary conductance due to the inelastic electron-boundary scattering, obtained using Eq. \ref{modelT} and the parameters given in the text, and
$\Sigma_{e-ph_2}^{tot}$ is the coupling coefficient when interpreting our results in terms of a lumped electron-phonon coupling per unit volume.
 }
\end{table}

Fig. \ref{electron_T}b analyzes small deviations from the $T^{4}$ dependence. In Fig.
\ref{electron_T}b we scale the measured electronic temperature $T_{e}$ by the heat power
dependent product $\left( P/\Sigma_{e-ph_2}^{tot} \right) ^{1/4}$. The obtained ratio
remains almost constant within two orders of magnitude of the applied power. The decrease in the data below 1 at small power $P < 0.4$ GW/m$^3$ is due to the crossover to the hot electron regime. A gradual small decrease is found at large Joule heating, which agrees with the theoretical curve derived from the two-temperatures-in-series model. The fitted curve, calculated using Eq. \ref{modelT}, displays a decrease that agrees well with the data.
This agreement indicates to us that the standard channel via overheated film phonons and Kapitza resistance plays a role in the heat transport to substrate phonons \footnote{The possible power flow limitation by $P_{e-ph_1} \propto T^5$
or $\propto T^6$ at the lowest temperatures  is not accessible in our experiment due to the large diffusive term $P_{diff}$ }.

We note that our results can be lumped  to a single electron-phonon coupling term of the form $\Sigma^{tot}_{e-ph_2} (T_e^4 - T_{ph_2}^4)$ per unit volume; the values for the effective coupling constant $\Sigma^{tot}_{e-ph_2}$ are given in Table II. The variation of this parameter could be assigned
to the different levels of disorder in each particular sample: e.g. the surface
roughness and/or quality of the nanostructure-substrate interface. Neglecting the Kapitza resistance, i.e. setting $T_{ph_1}=T_{ph_2}$, this kind of a description would  agree with the earlier electron-phonon coupling work on Ti films where the observed behavior was assigned to the disorder in the films \cite{Wei2008,Karasik2011}.

\section{Discussion}

From the comparison of the theoretical formulas and the measured data, it is evident that
an unambiguous identification of the  heat flow contributions in our Ti wire relaxation is
not possible. We can only provide a consistent picture based on the previous experimental works as
discussed in the description of the analysis of our results. Additional work would be needed to reach more definite
conclusions concerning the relevant terms and their relative magnitudes. In general,
the thermal resistance between the phonons in  metallic films and the substrate at low temperatures is
significantly smaller than the electron-phonon thermal resistance \cite%
{Roukes1985,Wellstood1989,Echternach1992}. However, this condition is
violated already in 100-nm copper films at electronic temperatures above $%
\sim 500$ mK \cite{Roukes1985,TaskinenPhD}. Hence, it is quite ordinary that, in a sample like ours,
the Kapitza resistance becomes relevant  in the experiments performed at $T_{e} \sim 1$ K.

 As already noted, we could optionally assign all our observations to heat relaxation by the electron-phonon scattering in the bulk.
 In this view, the electron-phonon term would be dominated by static impurities \cite{Sergeev2000} leading to a similar $\Sigma^{imp}_{e-ph_1}(T_e^4-T_{ph_1}^4)$ dependence as there exists for the Kapitza-limited  heat flow with $\frac{1}{4}A_K (T_{ph_1}^4-T_{ph_2}^4)$ per unit area.
 The effective coupling constant  would then become equal to $ 1/(d \Sigma^{imp}_{e-ph_1})+1/(A_K/4)$
 and there would be no way to separate between these two contributions from each other.
  Neglecting the Kapitza resistance and the term $P_{e-ph_2}$, we find $\Sigma^{imp}_{e-ph_1} =\Sigma^{tot}_{e-ph_2} = 1.5 \pm 0.5 \times 10^9$ Wm$^{-3}$K$^{-4}$, the magnitude of which is close to the values reported in Ref. \onlinecite{Wei2008}.

Our heat relaxation results are relevant also to
quasi-one-dimensional superconductors \cite{1Dsuperconductivity}. The issue of energy
dissipation in the resistive state of a superconductor is intimately linked with the
relaxation of the Joule heating of the same system in the normal state. Rapid
heat relaxation in narrow titanium nanowires in the normal state is essential for
obtaining single-quantum phase slip events in the superconducting state \cite%
{Lehtinen-R(T)-2012,Lehtinen-Bloch-2012}. According to our results, a reduction of the Kapitza
resistance between the film and the substrate could enhance the relaxation speed of the film and
thereby help in the nucleation of fully uncorrelated quantum phase slips.

In conclusion, our experimental analysis indicates that distinguishing the different heat flow contributions in narrow Ti wires is an intriguing issue and several assumptions have to be made in order to extract values for the separate contributions. Extensive additional experiments are needed in order to elaborate these issues further.

\section*{Acknowledgements}

This work was supported by the Academy of Finland (grant no. 284594, LTQ
CoE), and by the European Research Council (grant no. 670743). This research
made use of the OtaNano Low Temperature
Laboratory infrastructure of Aalto University, that is part of the European
Microkelvin Platform. Konstantin Arutyunov acknowledges the support of the
Russian Science Foundation grant No. 16-12-10521.


%

\end{document}